\documentclass[9pt,twocolumn,twoside]{pnas-mod}
\templatetype{pnasmodified} 
\usepackage{siunitx}

\newcommand{\fref}[1] {Figure~\ref{#1}}		

\newcommand{\T}{\top}										

\newcommand{\vect}[1]{\boldsymbol{#1}}                          
\newcommand{\matr}[1]{\mathbf{#1}}                              
\newcommand{\SetR}{\mathbb{R}}

\usepackage{soul}
\title{State-space aerodynamic model reveals high force control authority and predictability in  flapping flight}

\author[a]{Yagiz E. Bayiz}
\author[a,1]{Bo Cheng}

\affil[a]{Mechanical Engineering Department, The Pennsylvania State University, University Park, PA, 16801, USA}

\leadauthor{Bayiz} 

\authorcontributions{Author contributions: Y.B. and B.C. designed research; Y.B. performed research; Y.B. analyzed data; and Y.B. and B.C. wrote the paper.}
\authordeclaration{The authors declare no conflict of interest.}
\correspondingauthor{\textsuperscript{1}To whom correspondence should be addressed. E-mail: buc10@psu.edu}


\begin{abstract}
Flying animals resort to fast, large-degree-of-freedom motion of flapping wings, a key feature that distinguishes them from rotary or fixed-winged robotic fliers with limited motion of aerodynamic surfaces. However, flapping-wing aerodynamics are characterised by highly unsteady and three-dimensional flows difficult to model or control, and accurate aerodynamic force predictions often rely on expensive computational or experimental methods. Here, we developed a computationally efficient and data-driven state-space model to dynamically map wing kinematics to aerodynamic forces/moments. This model was trained and tested with a total of 548 different flapping-wing motions and surpassed the accuracy and generality of the existing quasi-steady models. This model used 12 states to capture the unsteady and nonlinear fluid effects pertinent to force generation without explicit information of fluid flows. We also provided a comprehensive assessment of the control authority of key wing kinematic variables and found that instantaneous aerodynamic forces/moments were largely predictable by the wing motion history within a half-stroke cycle. Furthermore, the angle of attack, normal acceleration, and pitching motion had the strongest effects on the aerodynamic force/moment generation. Our results show that flapping flight inherently offers high force control authority and predictability, which can be key to developing agile and stable aerial fliers.
\end{abstract} 

\dates{This manuscript was compiled on August 4, 2021}

\begin{document} 

\maketitle
\thispagestyle{firststyle}
\ifthenelse{\boolean{shortarticle}}{\ifthenelse{\boolean{singlecolumn}}{\abscontentformatted}{\abscontent}}{}

\section{Introduction}
Biological fliers with flapping wings are distinguished from rotary and fixed-winged robotic fliers in their ability to exploit the motion of aerodynamic surfaces with substantially larger magnitude and degrees of freedom (DoF) \cite{dudley2002biomechanics, fry2003aerodynamics, shyy2013introduction}. Therefore, they excel at flight control using versatile wing kinematics that modulate aerodynamic forces to a greater extent, thereby achieve diverse flight modes and unparalleled flight stability and agility \cite{dickinson2005initiation, cheng2011translational, sun2014insect, cheng2016flight, bomphrey2017smart, taha2020vibrational}. Not surprisingly, flapping flight has become an appealing design paradigm for aerial robotics, especially with the recent advancement in miniature electronics and fabrication techniques \cite{wood2008first, roll2015electromagnetic, jafferis2019untethered, zhang2017design, karasek2018tailless}.

Flapping-wing aerodynamics are characterised by highly unsteady and three-dimensional vortex flows that are difficult to model \cite{shyy2013introduction}, especially for the large variety of wing kinematics used in flapping flight. In particular, accurate predictions of aerodynamic forces often rely on high-fidelity and expensive computational or experimental methods \cite{liu1998numerical, dickinson1993unsteady, sun2002unsteady, young2009details, shyy2010recent, song2014three, bayiz2018real, bayiz2019experimental}, while existing low-fidelity models (e.g., quasi-steady models) \cite{minotti2002unsteady, sane2002aerodynamic, ansari2006non, berman2007energy, whitney2010aeromechanics, lee2016quasi} are unable to capture the unsteady and nonlinear aerodynamic mechanisms.

Nevertheless, animals can seemingly master the complex aerodynamics of flapping wings without becoming an expert in fluid dynamics. This likely results from the common paradigm of biological control that yields both predictability and controllability of locomotion \cite{wolpert1995internal, wolpert2000computational, shadmehr2012biological}: animals can acquire an internal model of the physical environment for predicting consequences of their actions and use this model together with sensory feedback for completing various motor control tasks. Therefore, it can be hypothesised that the flapping wings may permit high control authority and predictability of aerodynamic forces without relying on the explicit knowledge and direct control of the fluid flows, despite the complex vortex flows generated during the process. For predictability, this means that the kinematic history of the wings (i.e., the actions) should encode all necessary information to predict the forces generated by the flapping wings (with relatively small external flows), while the fluid dynamical system \emph{per se} does not impose any notable informational complexities to the aerodynamic forces experienced by the fliers. From the control authority, this implies that wing kinematic control inputs have strong and direct (i.e., linear and with small delay) influences on the aerodynamic forces and moments.

\begin{figure*}[!t]
\centering
\includegraphics[width=.95\linewidth]{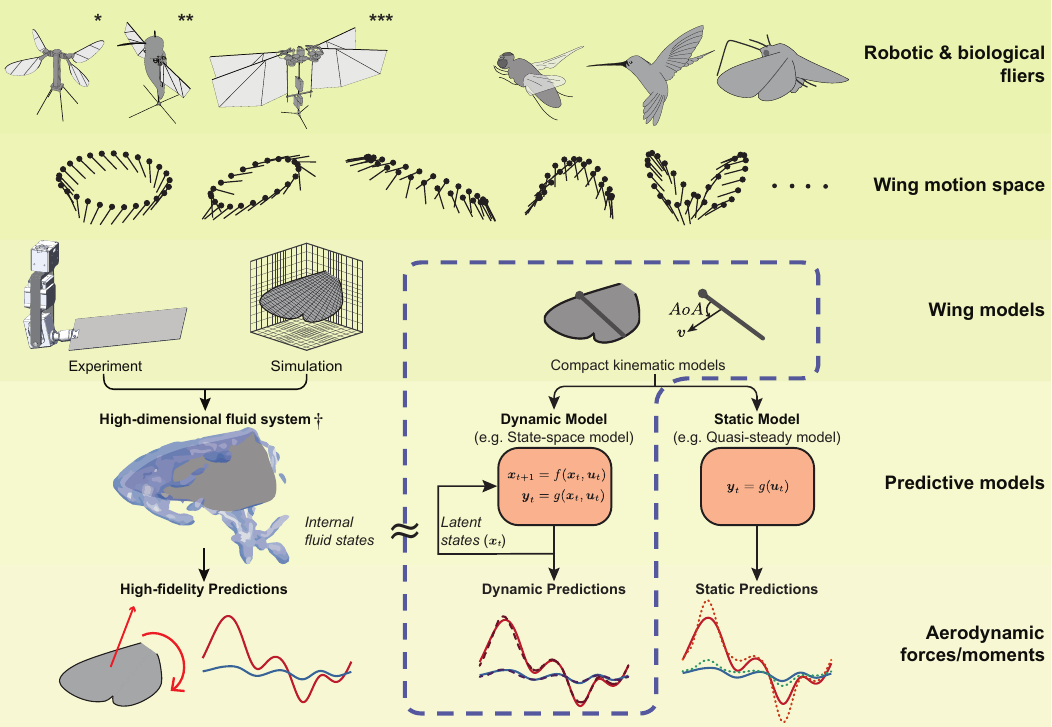}
\caption{\textbf{An overview of the predictive aerodynamic models.} Flapping fliers can leverage versatile wing kinematics from a large wing motion space for high flight control authority. The predictions of the forces and moments generated by the flapping wings often rely on expensive experimental or numerical simulation methods. The static models fail to capture the internal fluid states of the high-dimensional fluid system. This work provides a compact dynamical predictive model that can capture the latent internal states of the fluid system based only on the wing kinematics and force/moment data. The dashed box shows the proposed model. This model tries to capture the dynamic structures of the fluid system with its latent states. Thus, its force/moment predictions (dashed predictions in the last row) are more accurate than the static model predictions (dotted predictions in the last row). The illustrations of (*) Robobee \cite{jafferis2019untethered}; (**) Purdue Hummingbird Robot \cite{zhang2017design}; (***) Delfly Nimble \cite{karasek2018tailless}. ($\dagger$) Image is reconstructed from CFD analysis in \cite{lua2016aerodynamic}.}
\label{fig:fig1}
\end{figure*}

This work presents a predictive state-space model of flapping wing aerodynamic forces that was trained without explicit knowledge of the fluid flow. With as few as 12 states, this model can accurately predict the force and moment generated by 548 distinct wing motion kinematics, surpassing the quasi-steady (QS) models (\fref{fig:fig1}). This model implicitly captures the unsteady and nonlinear fluid effects underlying the force generation process of flapping wings, using a computationally efficient, data-driven nonlinear state-space model, namely Probabilistic Recurrent State-Space Model (PRSSM) \cite{doerr2018probabilistic}. Also, we further reveal the high control authority of key wing kinematic control variables on the force production and learned latent states via a cross-correlation analysis.

\section{Quasi-steady and state-space models of flapping flight aerodynamics}
QS models have been widely used to predict and analyse aerodynamic forces acting on flapping wings as they provide a computationally-efficient alternative to high fidelity models. QS models are direct static maps that convert local instantaneous wing kinematics to aerodynamic forces and moments; therefore, they do not require modelling of explicit flow features. These models have been developed using various aerodynamic theories, e.g., steady fixed-wing theory \cite{weis1973quick}, revolving wing theory \cite{ellington1996leading}, rotational circulation and added mass \cite{dickinson1999wing, sane2002aerodynamic}, and often use (semi-)empirical fitting of model parameters to force data generated by high fidelity computational or experimental methods. There have been many improvements to the accuracy of QS models (e.g., \cite{whitney2010aeromechanics, lee2016quasi, nakata2015cfd}), mostly from the one proposed by Sane and Dickinson \cite{sane2002aerodynamic}.

Notwithstanding their popularity, QS models are prone to several drawbacks that limit their use as an accurate and general predictive model for flapping flight to date. The majority of these models are empirically fitted to a small set of kinematics. Such models cannot be extended to a large variety of wing kinematic trajectories; thus, they only have a small range of validity \cite{nakata2015cfd, pohly2018quasi}. Notably, Nakata et al. proposed a data-driven approach \cite{nakata2015cfd} to fit a physics informed linear regression model to CFD data to improve local accuracy. Although this data-driven approach offers an excellent prediction for the region of interest, this approach has not been tested beyond the training set or wide range of wing kinematics. Additionally, the QS assumption ignores the effects of wing motion history or the internal flow states; thus, QS models can fail to capture all essential unsteady mechanisms, including wing-wake interaction \cite{sane2002aerodynamic, nakata2015cfd, pohly2018quasi}. Static models like QS models cannot easily capture the “dynamical system” underlying the aerodynamic force generation processes of flapping wings. Taha et al. \cite{taha2014state} partially mitigated this problem by introducing the Wagner effect \cite{dickinson1993unsteady, sane2003aerodynamics} to QS models. However, a general, compact, and computationally efficient dynamic mapping between the instantaneous forces and the wing motion is still missing in the literature.

State-space models (SSMs) \cite{durbin2012time} capture the system dynamics through latent states that hold enough information from the past to predict future states and measurements, thereby providing a compact representation of dynamic systems (e.g., two-dimensional viscous unsteady aerodynamics \cite{taha2021state}). However, identifying complex dynamic systems, such as flapping wing aerodynamics, poses a significant challenge for traditional (parametric) SSMs since a proper model structure is essential to train a compact model successfully. Gaussian Processes State Space Models (GPSSMs) \cite{wang2007gaussian, deisenroth2012expectation, frigola2013bayesian, frigola2014variational, svensson2016computationally, eleftheriadis2017identification} emerged as a versatile non-parametric nonlinear SSM, leveraging the excellent prediction capabilities and the probabilistic properties of Gaussian Processes (GPs) \cite{williams2006gaussian}. The GP approximation in SSM alleviates the a priori knowledge necessary for capturing the nonlinear behaviour while providing a measure of uncertainty for the predictions \cite{frigola2014variational}. PRSSM \cite{doerr2018probabilistic} is a type of GPSSMs and proved to be scalable and robust, making it ideal for large, complex problems.

\section{PRSSM as a predictive model of flapping wing aerodynamics} %
\begin{SCfigure*}[\sidecaptionrelwidth][ht]
\centering
\includegraphics[width=11.4cm]{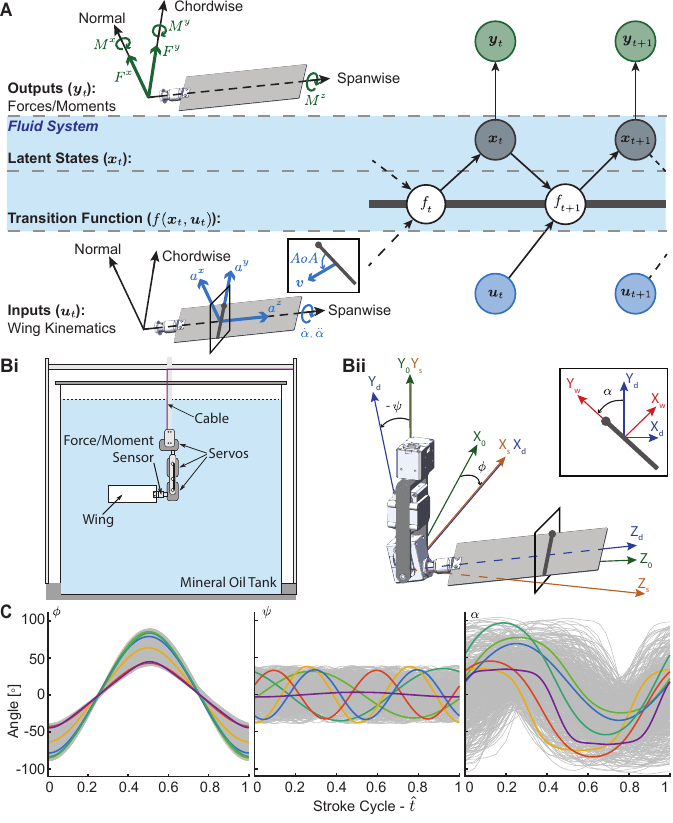}
\caption{\textbf{The illustration of the used methodology, robotic wing setup, and the collected data.} (\textbf{A}) Graphical illustration of the flapping wing aerodynamics with PRSSM. The model takes wing kinematics as inputs (blue nodes) and outputs aerodynamic forces and moments (green nodes). Both inputs and outputs are represented in the wing frame; thus, the model is coordinate frame independent. The latent states inherent in the fluid system are denoted with the grey nodes. White nodes indicate the values of the transition function, which are jointly Gaussian under a GP prior. (\textbf{B}) An overview of the experimental setup. (\textbf{Bi}) A schematic of the dynamically scaled robotic wing. The robotic is driven with three servo motors. The 6-axis force/torque sensor measures the aerodynamic forces and moments applied on the wing. The robotic wing was submerged into a mineral oil tank to maintain the Reynolds number at 1000. (\textbf{Bii}) Detailed illustration and the coordinate frames used in the robotic wing. The wing was designed to be rigid and flat, and it rotates around a pivot. The wing motion is described by three Euler angles, namely stroke angle  $\phi$, the wing position in the horizontal stroke plane, deviation angle $\psi$, the vertical wing deviation from the stroke plane, and the pitch angle $\alpha$. Coordinate frame subscripts  $0$ (global), $s$ (stroke), $d$ (deviation), $w$ (wing)  denote the corresponding frames. The angle definitions were only used to transcribe motion in the robotic wing and illustrating the wing trajectories. (\textbf{C}) The collection of all 548 trajectories was used to train and test our model. The coloured trajectories signify the test trajectories demonstrated in the results section. Blue: Test Trajectory $\#26$. Orange: Test Trajectory $\#14$. Yellow: Test Trajectory $\#13$. Purple: Test Trajectory $\#33$. Green: Test Trajectory $\#5$. Teal: Test Trajectory $\#32$.}
\label{fig:fig2}
\end{SCfigure*}

This work’s primary hypothesis is that a general and compact model can accurately predict the aerodynamic forces generated by flapping wings without relying on explicit knowledge of the fluid flow. Specifically, the aerodynamic forces generated by a broad range of wing motions can be accurately predicted with a direct, compact mapping from a set of wing kinematic variables. Unlike the quasi-steady models, the proposed method here uses a dynamic mapping rather than a static transformation. In particular, we assume that the flapping flight aerodynamics can be approximated by the following dynamic system,

\vspace{-9pt}
\begin{align*}
    \vect{x}_{k+1} &= f(\vect{x}_k,\vect{u}_k) + \vect{\omega}_k,\\ 
    \vect{y}_{k} &= g(\vect{x}_k,\vect{u}_k) + \vect{\epsilon}_k, \numberthis
\end{align*}

\noindent where $\vect{x}_k \in \SetR^{D_x}$  is the latent fluid state variable, which is a condensed representation of all the features of the fluid system underlying the aerodynamic force generation processes at time step $k \in [1, 2, 3,\hdots, K]$, $K$ being the terminal state, $\vect{u}_k \in \SetR^{D_u}$ denotes a known external input of wing kinematics, and $\vect{y}_k \in \SetR^{D_y}$ is the aerodynamic forces and moments (\fref{fig:fig2}A). The system and measurement noises are shown as $\vect{\omega}_k$ and $\vect{\epsilon}_k$, respectively. The transition function $f(\cdot)$ models the dynamic mapping from the current fluid states and kinematic inputs to the next fluid states, whereas  $g(\cdot)$ transforms the latent states to outputs (i.e., forces and moments). This dynamic system is represented and trained as a PRSSM model, as shown in \fref{fig:fig2}A. To capture the fluid dynamics, PRSSM uses a sparse GP \cite{quinonero2005unifying, snelson2006sparse} prior on $f(\cdot)$ and a linear $g(\vect{x}) = \matr{C} \, \vect{x}$. Here, we select $\matr{C} = [\matr{I}_{D_y} \: \: \matr{0}]$ to ensure the first $D_y$ elements of $\vect{x}$ correspond to $\vect{y}$. The variational inference method \cite{blei2017variational} is used to learn model dynamics. For further details, see Supplementary Materials.

The set of wing kinematics $\vect{u}_k$ and aerodynamic forces $\vect{y}_k$ used in the model are represented in the wing frame, therefore independent of the coordinate frames or the Euler angles that describe the wing trajectories. In particular, $\vect{x}$, $\vect{u}$ and $\vect{y}$ are defined as,

\vspace{-9pt}
\begin{align*}
    &\vect{x} = [x^1, x^2, \dots, x^{D_x}]^\T,\\
    &\vect{u} = [|\vect{v}|, AoA, a^{x}, a^{y}, a^{z}, \dot{\alpha}, \ddot{\alpha}]^\T,\\
    &\vect{y} = [F^{x}, F^{y}, M^{x}, M^{y}, M^{z}]^\T, \numberthis
\end{align*}

\noindent where $x^i$  denotes the $i^{th}$ element of the latent state, $\vect{v}$ is the linear velocity of the wing at the unit distance along the axis of rotation from the wing’s base, and  $AoA$ is the geometric angle of attack. Furthermore, $a^x$ (normal), $a^y$ (chordwise), and $a^z$ (spanwise) represent the linear acceleration components of the wing at the unit distance along the axis of rotation from the wing’s base. Among those, $a^x$ and $a^y$ are due to angular acceleration, and $a^{z}$ corresponds to centripetal acceleration  (\fref{fig:fig2}A). The pitching velocity and the acceleration are denoted $\dot{\alpha}$ and $\ddot{\alpha}$ respectively. $F^x$ and $F^y$ are the normal and chordwise aerodynamic forces, respectively. Similarly, $M^x$, $M^y$, and $M^z$ denote the normal, chordwise, and spanwise aerodynamic moments applied on the wing. Note that the spanwise force $F^z$ is omitted in this work since the axial aerodynamic force in flapping wings is negligible.

The model was trained using data acquired from a dynamically scaled robotic wing \cite{bayiz2018real, chen2018unsteady, bayiz2019experimental}  (\fref{fig:fig2}B). In this work, we aimed to develop a general predictive model valid for a broad range of different wing motion kinematics while providing a novel and fundamental perspective towards flapping flight\textemdash characterized by fast, reciprocal, and 3DoF motion of the aerodynamic surface. We assembled a large dataset consisting of 548 randomly generated trajectories (\fref{fig:fig2}C). This dataset consisted of a wide range of flapping wing kinematics, including both ``naturalistic'' wing kinematics commonly observed in hovering and more arbitrary wing trajectories that can be possibly used in maneuvers.

The latent state dimension $D_x$ was determined through model accuracy. Briefly, we trained the PRSSM model with $D_x = 8, 12, 16$, and compared the accuracy of these models. We observed that the modelling performance noticeably improved when $D_x$ increased from $8$ to $12$, yet the accuracy of models with $D_x = 12$ and $D_x =16$ was insignificant. Thus, we selected $D_x =12$. Similarly,  we tested the effect of the size of the training data on training and testing accuracy of the model. To avoid over-fitting in our model, we identified the data size that resulted in similar the training and testing accuracy and partitioned our trajectories as 512 training and 36 test trajectories. For the details of both selection methodologies, see Supplementary Materials.

For the sake of comparison between static and dynamic mappings, we also trained a linear regression quasi-steady model (LRQSM), similar to the one proposed by Nakata et al. \cite{nakata2015cfd}. The instantaneous aerodynamic forces and moments were approximated as a linear mapping of kinematic features that have been previously identified \cite{sane2002aerodynamic, minotti2002unsteady, berman2007energy, whitney2010aeromechanics, nakata2015cfd}. These kinematic features were nonlinear combinations of the wing kinematics represented in the wing frame, similar to what we used for PRSSM inputs. Particularly, the aerodynamic forces and moments were approximated as,
\begin{equation}
    \vect{y} = \matr{A}_{QS} \; k_f(\cdot),
\end{equation}
where,
\begin{align*}
    \label{eq:kinematicfeatures}
    k_f(\cdot) = [&(|\vect{v}|)^2, AoA|\vect{v}|, AoA(|\vect{v}|)^2, AoA(|\vect{v}|)^3, \dot{\alpha}|\vect{v}|, \numberthis\\ 
    &\dot{\alpha}|\dot{\alpha}|, a^x, a^y, \dot{\alpha}v^x, \dot{\alpha} v^y ]^\T.
\end{align*}

\noindent It should be emphasized that the selected kinematic features can accurately estimate all the components (e.g. $\cos(AoA)|\vect{v}|^2$) of the QS models when linearly combined. Clearly, the transformation matrix $\matr{A}_{QS}$ can be determined using any form of linear regression, and here, LASSO regression \cite{friedman2001elements} is used. This method is the paramount example of a compact QS model for a given data set as it minimizes the error between the training data and predictions. Notably, a local linear regression model would probably provide better estimations as demonstrated in \cite{zheng2020kinematic}. However, such a method requires access to the whole data set to predict the aerodynamics at every single time instant. Accessing a large data set is not acceptable for a compact representation, thus disregarded in this work.

\begin{figure*}[!t]
\centering
\includegraphics[width=.95\linewidth]{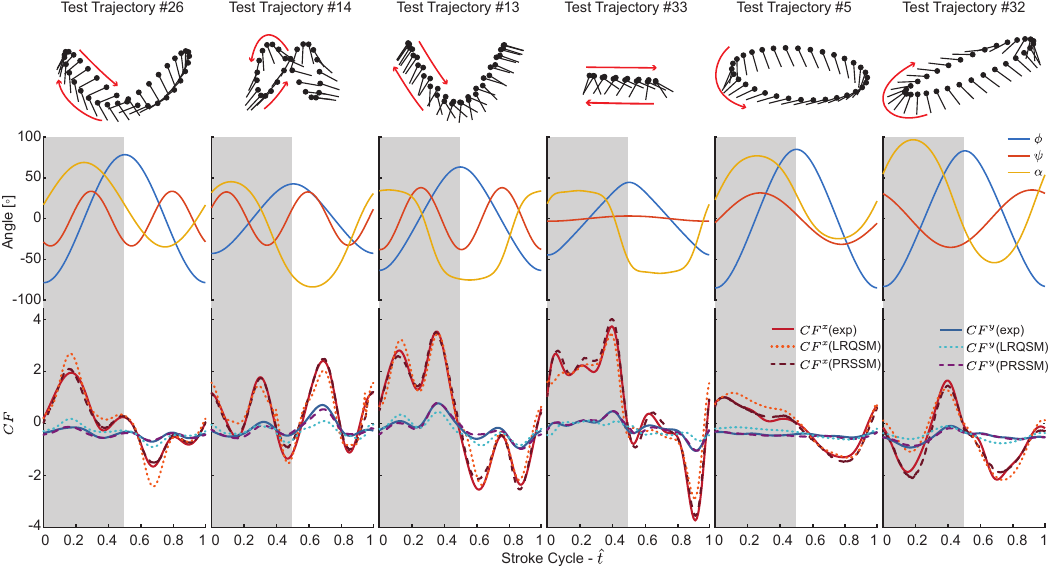}
\caption{\textbf{Time history of wing kinematics and the force coefficients.} $CF^x$ and $CF^y$, for six selected test trajectories. The forces $F^x$ and $F^y$ were scaled with the factor $1/2 \rho S U_{ref}^2$ to achieve $CF^x$ and $CF^y$, where $\rho$ denotes the fluid density, $S$ is the wing area, $\bar{c}$ is the chord length, and $U_{ref}$ is the average wing tip velocity. The trajectories are picked to demonstrate the versatility of the PRSSM representing the distinct and characteristic wing kinematics. Test trajectories $\#26$ and $\#14$ represent different trajectories with the figure of eight shape. Test trajectory $\#13$  is a U-shaped trajectory, whereas $\#33$ is flat. $\#5$ and $\#32$ are examples of oval-shaped trajectories. Every trajectory has a distinct angle of attack progression. Top row: Projections of the 3D kinematics of a chord slice on a 2D plane. The red arrow indicates the flapping direction. Middle row: Instantaneous wing kinematics. Bottom Row: Measured and predicted instantaneous force coefficients. Red-hued (red, orange, maroon) plots indicate normal force coefficients $CF^x$, blue hued (blue, teal, purple) plots chordwise force coefficients $CF^y$. The RMSE of the normal force coefficient ($CF^x$) for each trajectory (PRSSM/LRQSM): \#26 (0.11/0.33), \#14 (0.19/0.33), \#13 (0.17/0.35), \#33 (0.21/0.54), \#5 (0.13/0.26), \#32 (0.20/0.38). The RMSE of the chordwise force coefficient ($CF^y$), for each trajectory (PRSSM/LRQSM): \#26 (0.042/0.16), \#14 (0.094/0.271), \#13 (0.086/0.208), \#33 (0.063/0.240), \#5 (0.051/0.150), \#32 (0.04/0.190).}
\label{fig:result1}
\end{figure*}

\section{Results}
Our results show that dynamic mapping using PRSSM has a higher predictive power in terms of accuracy and generality than the static mapping using LRQSM for the flapping flight aerodynamics. We recognise that the LRQSM captured the general trend of the aerodynamic forces and moments, and it was shown to possess sufficient predictive power to serve as a fast and simple alternative. However, for the 512 training and 36 test trajectories, PRSSM yielded significantly higher prediction accuracy statistically for both instantaneous and cycle-averaged forces and moments. Furthermore, cross-correlation analysis based on the trained PRSSM revealed the strength and the delay of influences that these wing kinematic variables (e.g., $AoA$, $\vect{v}$, $a^x$, $\dot{\alpha}$ etc.) have on the resultant aerodynamic forces, moments, and latent states, i.e., the control authorities of different wing kinematic variables. Finally, PRSSM was computationally efficient; once trained, this model could predict forces and moments for each trajectory in less than 0.25 seconds (with a workstation-GeForce GTX1070 GPU, Intel 7700K CPU- and Tensorflow). Note that this computation speed still does not match those possible in small flying animals or directly applicable with onboard computation in small micro aerial vehicles, but it can be further improved with code optimization or using a different SSM.

\subsection{PRSSM outperforms LRQSM in predicting the instantaneous aerodynamics - Examples}
To demonstrate the prediction capability of PRSSM over a wide range of different wing kinematics, we present the time history of the aerodynamic forces for six test trajectories with distinct flapping patterns (\fref{fig:result1}). The force and moment predictions for all test trajectories can be found in the Supplementary Materials. The PRSSM force predictions in both normal and chordwise directions were in close agreement with the experimental data for all selected trajectories. On the other hand, LRQSM struggled to predict both force components accurately. For the normal component, LRQSM predicted the general shape and magnitude of the force profile, but it failed to consistently capture the correct magnitude of the peaks and sometimes was unable to predict their existence. Compared to PRSSM, the root mean square error (RMSE) of LRQSM normal force predictions were 1.7 to 3 times higher for the presented trajectories. For the chordwise component, LRQSM prediction accuracy was even worse, resulting in 2.5 to 4 times higher RMSE than PRSSM’s for the presented trajectories.

A closer examination and comparison of PRSSM and LRQSM predictions further suggests that PRSSM successfully captured the unsteady and nonlinear aerodynamic effects that were not included in quasi-steady models. For instance, trajectory \#26 (column 1) had a trajectory profile that was mostly sinusoidal and can be labelled as a classical figure-of-eight flapping trajectory (the $AoA$ was maintained mostly between \ang{0} and \ang{90}). For this trajectory, LRQSM failed to identify the magnitude of both normal force peaks around $0.15$ stroke cycles after the stroke reversal, where wing-wake interaction was likely to be still prevalent \cite{chin2016flapping}. We observed a similar behaviour at test trajectory \#33 (column 4), which had a classical flat flapping pattern with a flat stroke plane and a traditional AoA progression similar to what was studied by Sane and Dickinson \cite{sane2002aerodynamic}. At the start of the stroke ($\hat{t} \in [0, 0.15]$), a high peak that was likely associated with the wake capture mechanism \cite{dickinson1999wing, sane2002aerodynamic, birch2003influence} appeared, indicating a strong dependence on the history of wing motion. PRSSM successfully predicted this peak with its dynamical mapping capability, while the LRQSM, which does not model any history-dependency, failed to capture the peak as expected.

Moreover, for trajectories \#5 (column 5) and \#32 (column 6), the LRQSM failed to predict the normal force for both trajectories during the majority of the first half stroke  ($\hat{t} \in [0, 0.5]$) These trajectories had unconventional $AoA$ profiles in the first half strokes - close to zero for trajectory \#5 and mostly negative for trajectory \#32. Such trajectories are well understudied as they do not commonly appear in the steady flight of biological fliers; thus, they are usually ignored in QS models. As a result, LRQSM kinematic features (eqn. \ref{eq:kinematicfeatures}) might have failed to capture such trajectories’ aerodynamics, indicating a possible limitation on its range of validity. In comparison, PRSSM relies on GP inference to implement the nonlinearity, therefore, accurately predicting these force profiles.

\subsection{PRSSM achieves high accuracy and generality - Statistics of prediction accuracy}
\begin{figure*}[!t]
\centering
\includegraphics[width=.95\linewidth]{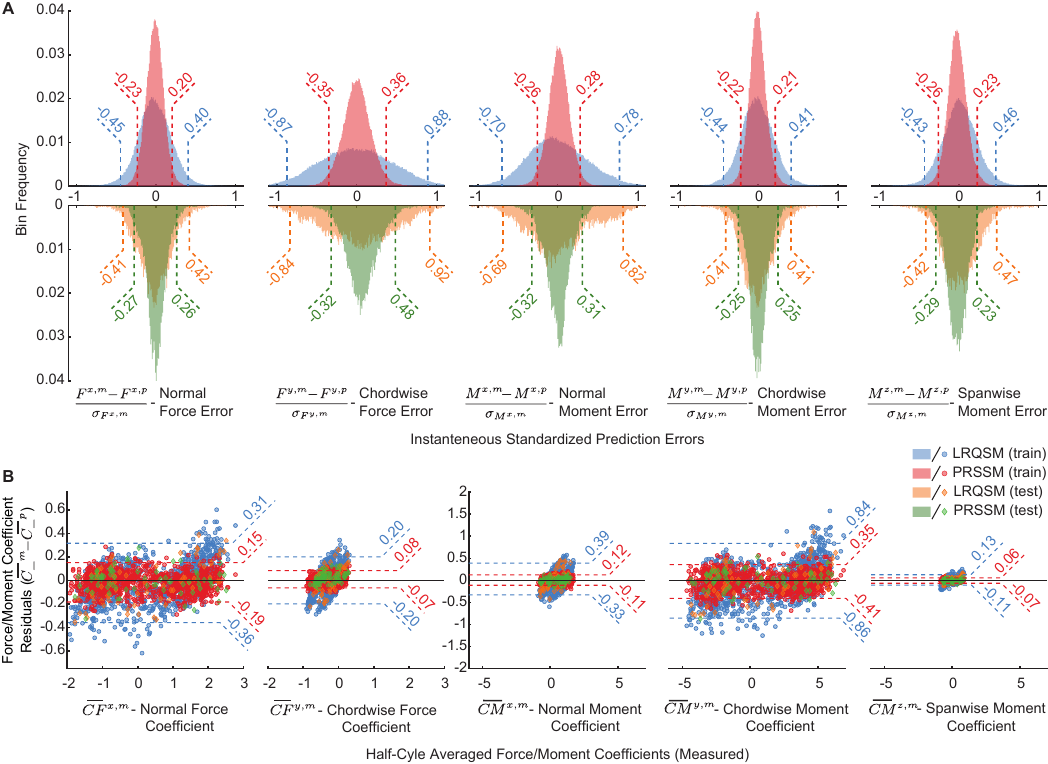}
\caption{
\textbf{The prediction accuracy statistics of the PRSSM and LRQSM.} The superscript $m$ indicates measured data points and $p$ indicates the predicted data points. The dashed lines shows the $\%95$ confidence bounds for the corresponding data. (\textbf{A}) The histograms of the standardised error for each aerodynamic force and moment component. The prediction error for each output was standardized with the standard deviation $\sigma$ of the training data points for the corresponding output variable which is indicated in the subscript. Numbers on the dashed lines are the confidence bounds of the standardised error and corresponds to the data set with matching the colour. The histograms were plotted with 200 bins. (\textbf{B}) Cycle averaged measured force/moment coefficients versus the residual the cycle averaged measured force/moment coefficients. The coefficients in this figure are averaged over the half-cycle starting from each stroke reversal, which is indicated by the ``~ $\bar{ }$ ~" operator. Similar to $CF^x$  and $CF^y$ in \fref{fig:result1}, the moments $M^x$, $M^y$, and $M^z$ were scaled with $1/2 \rho \bar{c} S U_{ref}^2$ to calculate $CM^x$, $CM^y$, and $CM^z$. Numbers on the dashed lines are the confidence bounds of the cycle averaged prediction errors. Note that for this figure, the confidence intervals are calculated with training and test data together and represented with the training data colour.}
\label{fig:result2}
\end{figure*}

Here we further illustrate and compare the instantaneous and cycle-averaged prediction accuracy of the PRSSM and LRQSM. First, the histograms (i.e., frequency distribution) and the $95\%$ confidence intervals (CI) of the standardised errors of instantaneous force and moment predictions are shown in \fref{fig:result2}A. Here, the error for each output variable was standardised by the standard deviation ($\sigma$) of the entire measurement data set for the corresponding output. Therefore, here confidence bound equal to 1 means that the prediction error can reach one $\sigma$ of the force (or moment) distribution of the entire trajectory set. The error distributions of both PRSSM and LRQSM for all force and moment components were mostly unimodal and symmetrical around 0. We quantified the prediction errors using the spread of the error measured by $95\%$ CIs. The PRSSM errors were narrowly concentrated around 0, while those of LRQSM were significantly more dispersed. For all the force or moment components, the confidence bounds of LRQSM (0.40-0.88 for the training set, 0.41-0.92 for the test set) were approximately two to three times higher than those of PRSSM (0.20-0.36 for the training set, 0.23-0.48 for the test set). Therefore, the expected instantaneous error from the PRSSM prediction was less than half of what was expected from LRQSM.

These results further showed the superiority of the dynamic mapping to accurately model the flapping flight aerodynamics and the shortcomings of the static mapping in the LRQSM model to provide a proper structure for the flapping flight aerodynamics. Note that the variability for the chordwise force component $F^y$ was greater than that of the normal component $F^x$ for both PRSSM and LRQSM, which was expected considering the limitations of our experimental setup. The normal force had a significantly larger magnitude; thus, the chordwise force measurement’s signal to noise ratio was lower given our force/moment sensor’s resolution and sensitivity (see Supplementary Materials).
 
Next, we investigated the models’ cycle-averaged performance, as the errors between the measured and predicted cycle averaged force/moment coefficients are shown in \fref{fig:result2}B. Compared to the LRQSM, the cycle-averaged prediction errors of PRSSM were again more narrowly concentrated around the zero error line. The amount of spread was consistent with our instantaneous error analysis as the PRSSM CIs for combined training and testing data sets are about two to three times more concentrated than those of LRSSM. Specifically, the PRSSM $95\%$ CIs were between $[-0.19, 0.15]$ for the normal force coefficient and $[-0.07, 0.08]$ for the chordwise coefficient. Comparatively, the same prediction errors for LRQSM were $[-0.36, 0.31]$ and $[-0.20, 0.20]$ and $[-0.20, 0.20]$, respectively. Furthermore, LRQSM prediction errors showed some increasing trends with the magnitude of the force/moment coefficients, as the PRSSM prediction errors remained more invariant across the different magnitude of the coefficients. A notable example of those trends was the linear trend in $\bar{CF}^y$ distribution, further exemplifying that LRQSM cannot capture certain structures in the flapping flight aerodynamics.

\begin{figure*}[!t]
\centering
\includegraphics[width=.95\linewidth]{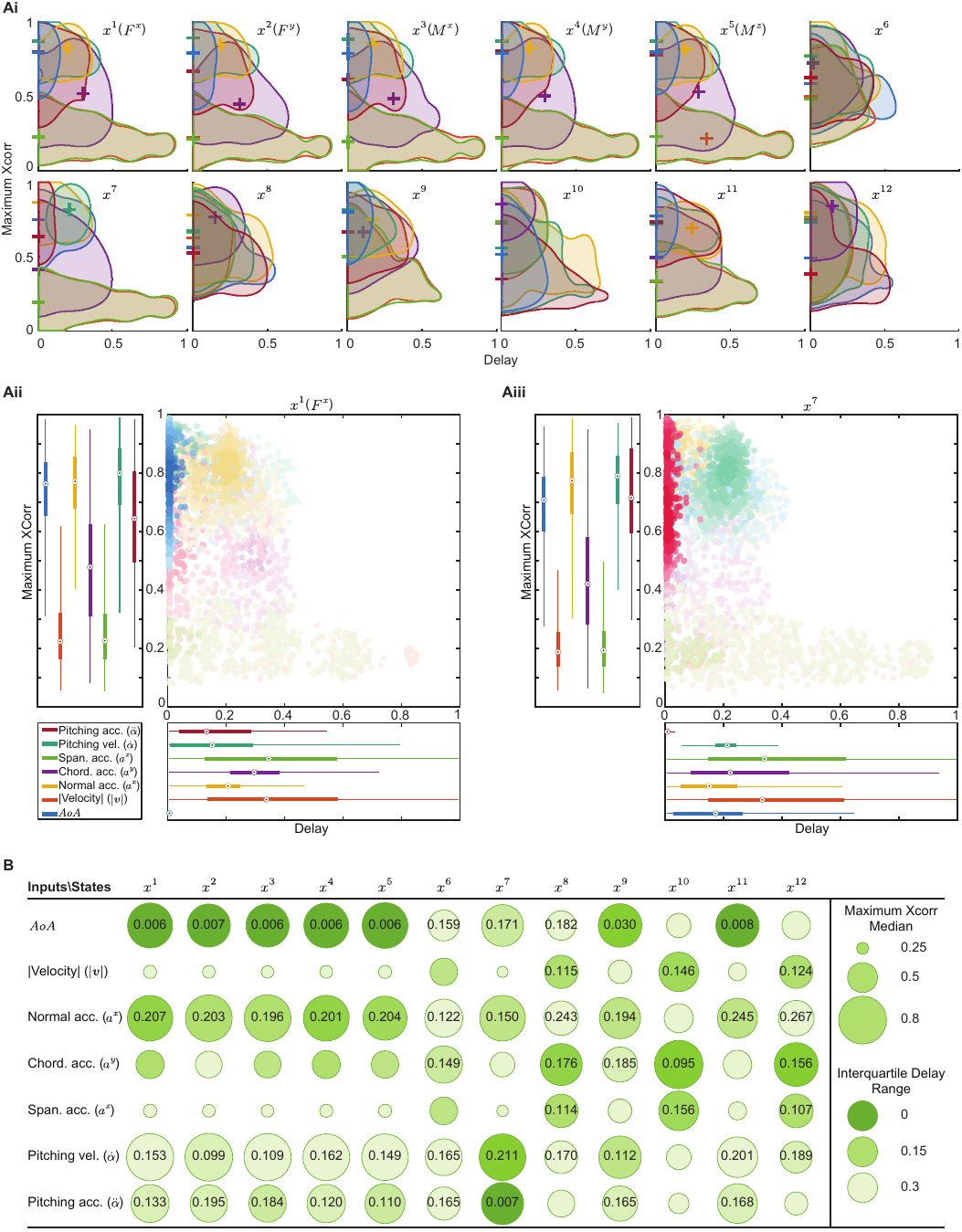}
\caption{
\textbf{Cross-correlation analysis between the latent states ($\vect{x}$) and the kinematic input variables ($\vect{u}$).} (\textbf{A}) The distribution of the maximum absolute cross-correlation - time delay pairs. Colourmap is the same for all figures. (\textbf{Ai}) Overview of all distributions. Each shaded area represents the $70\%$ confidence region with the highest likelihood, and the (+) sign represents the maximum likelihood point of the distribution of the corresponding kinematic variable. The distributions were reconstructed using kernel probability density estimation \cite{bowman1997applied}. (\textbf{Aii, Aiii}) Scatter plots of the cross-correlation results are shown for the state $x^1$, the one corresponding to the normal force (\textbf{Aii}), and the state $x^7$ (\textbf{Aiii}). The lightness of the colour in each point indicates the likelihood of the corresponding data point obtained from reconstructed distributions, and darker colours imply a higher likelihood. The circles in the box plots are the median, the thick line denotes the interquartile range, and the thin lines indicate the \%95 CI. (\textbf{B}) Summary of cross-correlation analysis. Circle size indicates the median of the maximum cross-correlation between each state-input pair; the larger the radius higher the correlation. The lightness indicates the interquartile range of the corresponding delays; darker colours imply higher clustering. This signifies how dispersed the delay with the highest cross-correlations among different experiments. The numbers inside the circles indicate the median of the delays for pairs with high cross-correlations (median cross-correlation larger than $0.5$).}
\label{fig:result3}
\end{figure*}\textbf{}

\subsection{Wing kinematic variables have different control authorities on states: cross-correlation analysis} The latent states of the trained PRSSM encapsulate the aerodynamic force generation process of flapping wings. Therefore, they can reveal innate dynamical system behaviours that are not directly observable, potentially through a cross-correlation \cite{box2015time} analysis of the latent states. Specifically, the cross-correlation provides a metric for the linear similarity of two variables with different forward or backward time delays. Thus, it can reveal both the strength and the delay of the influences exerted by the wing kinematic inputs on the states, including the aerodynamic forces and moments produced.

Here, we calculated the cross-correlation between each latent state variable $\vect{x}$ and the input $\vect{u}$ for each training and test trajectory. To maintain the causality between inputs and outputs, we limited our search only to the positive delay region where the states $\vect{x}$ had a time delay from the inputs $\vect{u}$. Furthermore, we normalised the time delays with the stroke cycle duration and focused on the correlation within a full stroke cycle. Once we calculated cross-correlations, we identified the amount of delay with the highest absolute cross-correlation. Note that the model we presented here was nonlinear; thus, the cross-correlation can be potentially blind to some nonlinearities.

The cross-correlation analysis results are summarised in \fref{fig:result3}, with \fref{fig:result3}Ai showing the overall distributions of the highest absolute cross-correlation and the corresponding lag among different trajectories for each state (total of 12 states). At first glance, the cross-correlation distributions’ upper-right region was empty, indicating that the high cross-correlation input-state pairs were exclusively clustered within zero to half stroke cycle delay. A closer inspection of the actual scattering of the data points (e.g. \fref{fig:result3}Aii for $x^1$ and \fref{fig:result3}Aiii for $x^7$-rest of scatter plots can be found in Supplementary Materials) showed that the data points with strong correlations were consistently clustered on the right side of the plot (delay $\in [0, 0.5]$). Such clustering suggests a short but non-zero history dependency (likely to be shorter than half a stroke cycle) of the states on the wing kinematic inputs.

The distributions of the five output states (first five states of \fref{fig:result3}) are similar to each other, where pitching-motion related inputs ($AoA$, $\dot{\alpha}$, and $\ddot{\alpha}$) and the normal acceleration of the wing $a^x$ had very strong correlations with the aerodynamic forces and moments with varying delays. The tabular graphic in \fref{fig:result3}B  confirms this similarity among the distributions of the output states (first five columns), as the correlation strength (size of the circle), delay median locations (numerical values within the circle), and the corresponding dispersion of the delays (lightness of the circle) are almost identical for the output states. Since these forces and moments were resulted from integrating the same pressure distributions over the wing surface in different directions and with moment arms, such similarity suggests that the integration process does not alter the linear dependency on the kinematic input variables, i.e., each kinematic variable has similar control authority over all of the force and moment components. 

Notably, the correlations between the $AoA$ and the output states were consistently distributed around 0 delay and correlation value 0.75, as seen from \fref{fig:result3}Ai and \fref{fig:result3}B. This indicates a linear and instantaneous influence of $AoA$ over the aerodynamic forces and moments, suggesting its high control authority with consistently low history dependency. Other components of the pitching motion ($\dot{\alpha}$ and $\ddot{\alpha}$) also had strong linear influences on the outputs as their maximum cross-correlation values are consistently high (medians within $[0.7, 0.8]$). However, corresponding delay distributions were scattered and did not point to a distinct delay mechanism, suggesting high control authority but with trajectory-specific history dependencies. In contrast, the normal acceleration had strong similarities with the force and moment components with relatively clustered delay locations. Although it was not as compact as $AoA$, the optimal delay distribution for $a^x$ occurred persistently around $20\%$ of the stroke cycle duration. The scatter plot \fref{fig:result3}Aii validates these conclusions for the normal force. Together, these suggest a high control authority of normal acceleration but with consistently high history dependency, possibly indicating circulatory-based forces (with delay) are more dominant than added-mass \cite{sane2002aerodynamic} forces (no delay) among the training and testing trajectories.

Unlike the output states (forces/moments), the linear relationships between the kinematic inputs and the latent states($x^{6}$-$x^{12}$) show state-specific, distinct behaviours (with different correlation strength, delay amount, and dispersion). Each latent state possibly captured a unique system behaviour inherent to the force/moment production processes, which cannot be directly observed from forces/moments. Among these latent states, $x^{7}$ had several noteworthy features in its cross-correlation distribution (\fref{fig:result3}Ai second row, second column and \fref{fig:result3}Aiii). For this state, the clustering of the pitching acceleration $\ddot{\alpha}$ was exclusive to the 0 delay region with a high maximum cross-correlation coefficient (median $0.72$), resembling those of $AoA$ to output states described above. This is indicative of a strong instantaneous influence of $\ddot{\alpha}$ over the latent state $x^7$. Additionally, pitching velocity $\dot{\alpha}$ also had a prevalent linear impact on the state $x^7$ as the median maximum cross-correlation is  $0.80$  with a tight clustering. In contrast to $\ddot{\alpha}$, the corresponding delay for $\dot{\alpha}$  was at  $21.1\%$ of the stroke cycle, implying a hidden and delayed force/moment generation mechanism related to $\dot{\alpha}$. The other notable linear correlations for latent states, such as the delayed influence of the chordwise acceleration on  $x^8$ (delayed by $17\%$ of the stroke cycle), $x^{10}$ ($9.5\%$ of the stroke cycle), and $x^{12}$ ($15.3 \%$ of the stroke cycle), can be identified from \fref{fig:result3} and the scatter plots in the Supplementary Materials.

\section{Discussion}
In this work, we developed a compact, state-space model of flapping wing aerodynamics that can accurately predict aerodynamic forces and moments for a wide range of wing kinematics without explicit knowledge of fluid flow. This model’s existence reveals the inherent high predictability and high-control authority in flapping wing aerodynamics. These are likely due to the persistent excitation of the fluid system by the continuous, large-DoF flapping wing motion. Therefore, the quick-moving aerodynamic surfaces in engineering design might be key to developing agile, stable, and robust robotic fliers. 

\subsection{Unsteady aerodynamics may pose a lesser problem to flight control of flapping fliers}
Our results imply that an intelligent flyer (robotic or biological) can learn an accurate predictive model of the flapping flight directly through wing kinematics. This can be likely attributed to the fast and large-DoF flapping wing motion that persistently excites the fluid flow and dominates the information required to predict the aerodynamic force and moment, giving rise to the high control authority in flapping flight. In comparison, for classical flow or turbulence control problems in fixed or rotary airfoils \cite{wu1998post, amitay2001aerodynamic,brunton2015closed}, forcing or change in the airfoil configuration applied on the wing is limited. Therefore, the amount of excitation and control on the fluid system, consequently, the control over aerodynamic forces may be substantially lower than those of the flapping wings. As a result, the resultant forces become more directly influenced by innate fluid dynamical behaviours.

Furthermore, the fluid dynamical system possibly does not impose any significant informational complexities on the aerodynamic forces for flapping wings. The information needed to predict the flapping aerodynamic forces accurately is well encoded in the wings’ kinematic history. Thus, the flapping flight inherently has high predictability of aerodynamic forces/moments from the wing motion. This further implies that, despite the complex fluid flow behaviours of flapping wings, it is theoretically less challenging for a flying animal or an intelligent robot to predict the forces or any concomitant sensory cues resulting from their wing motion (i.e., the flight control inputs). Therefore, the unsteadiness of the flapping aerodynamics may pose a lesser problem to fliers’ control systems than originally postulated.

These two conclusions lead to a few notable implications on the control paradigm of flapping fliers, including an internal forward model of the physical environment that enables the flier to predict the consequences of its actions for motor control purposes. An internal model predicting the forces or moments acting on the wing and subsequently on the body (or the concomitant effects of these forces and moments on their sensory cues) can be propitious for both reflexive feedback control strategies as well as the feedforward pathways \cite{wolpert1995internal, wolpert2000computational, shadmehr2012biological}. With our model, we show that, for the locomotion control problem, robot or animal alike, the wing motion and the resulting aerodynamic forces are sufficient for learning an accurate, compact, and general internal model without the explicit knowledge of the fluid flow field. For a flier, establishing a predictive model of aerodynamics requires only the causal relationship between the motor actuation and the pertinent force output measured in a certain form by their sensory systems, not the fluid’s behaviour \emph{per se}.

For a biological flier, such a model may exist between the muscle activation and the proprioceptive senses, potentially in the form of physiological internal models observed in biological motor control (i.e., central pattern generators \cite{ijspeert2008central}, efference copies \cite{manoonpong2013neural}). Considering the sensory feedback available to biological fliers, our results imply that if such a model exists, it does not necessarily need to include any sensory measurements of the explicit flow field features, such as the vortex locations. Together, the high control authority of the flapping wings, combined with predictive and feedback control, are likely to be two of the key features leading to the high stability, manoeuvrability, and robustness to unexpected external perturbations observed in natural flapping-wing fliers. Undoubtedly, for biological fliers, the control authority and predictability ultimately depend on the coupled wing-motor system (e.g., thorax for insects) and the morphological and structural properties of the wings. For example, mechanisms such as wing/feather deformation, passive wing pitching, and muscular control can be beneficial or detrimental to the fliers' ability to harness control authority intrinsic to the flapping flight.

From the perspective of aerial robotics, designing a wing-actuation system that truly leverages the intrinsic advantages of flapping flight instead of merely emulating its simple flapping-form might be critical for eventually developing robotic fliers with higher agility and stability than conventional designs with fixed or rotary wings. The ability to predict the aerodynamic forces solely from wing motion also offers the flier an array of potent control strategies. Including the full body and actuator dynamics, one can train an accurate model from actuator signals to body kinematics, provided that the flier can measure both. Therefore, the existence of such a predictive model, together with the high control authority of the flapping wing, renders the classical flow control problem in airfoils seemingly unnecessary for locomotion control of flapping-wing fliers.

\subsection{Aerodynamic states mainly depend on the wing motions in less than a half cycle}
The cross-correlation analysis of the latent states is the second major contribution of this work. This analysis revealed that the latent state-wing kinematics pairs are strongly correlated only when the delay is less than half a flapping cycle (\fref{fig:result3}). Such a clustering indicates a short but non-zero history dependency (i.e., likely to be shorter than half a stroke cycle) of the states on the wing kinematics. In other words, the aerodynamic forces and moments can be accurately predicted, given that the wing kinematics history for half cycle is provided. Intriguingly, this is consistent with the minimum-domain impulse theory for unsteady aerodynamic forces, recently proposed by Kang et al. \cite{kang2018minimum}. Briefly, this theory states that the fluid forces acting a body (e.g., a wing) depend exclusively on, therefore can be predicted entirely by the states of the vortical structures still connecting to the body. It is known that the majority of the vortex shedding occurs at the stroke reversals (i.e., on each half wing cycle) \cite{birch2003influence, cheng2014three}, and the vortical structures attached to the wings are almost exclusively generated within a half stroke. Therefore, this theory supports our conclusion that the flapping flight aerodynamic forces/moments should mainly depend on the wing motions in less than a half cycle.

\subsection{Angle of attack has an immediate effect on forces and moments}
We can infer the control authority that each wing kinematic variable has over the aerodynamic force and moment generation. For instance, $AoA$  has a strong, immediate linear effect on the forces and moments. Thus, not only that the fliers can use it to generate immediate modulation of aerodynamic forces, but they should also be able to predict the outcomes of this manipulation more easily compared to other inputs. Together, these features may imply $AoA$ as a favourable control input by the flight control system. As another example, we showed that the pitching acceleration $\ddot{\alpha}$ has a strong linear and immediate influence over one of the latent states ($x^7$) instead of the forces and moments, so its effects will not be directly observable to the flier. The observed contribution of the pitching acceleration on the aerodynamic forces and moments does not have a consistent delay amount (\fref{fig:result3}); thus, it is harder for the flier to predict.

Furthermore, the strong linear correlations for both output and latent states point out the timing and location of the possible aerodynamic mechanisms occurring in flapping flight. For instance, the delayed effects of the normal acceleration that we observed here might indicate a novel aerodynamic mechanism contributing to the force generation in flapping flight or change our understanding of an existing mechanism. With these insights, we can advance our understanding of flapping flight locomotion control, design better controllers for robotic fliers, and develop even simpler (potentially static) aerodynamic models.

\acknow{We thank Dr. Jesse Roll, Benjamin Cellini, Yu Pan, Dr. Jean-Michel Mongeau, and Dr. Haibo Dong for their valuable comments and support on this work.}

\funding{This research is funded by Army Research Office (grant\# W911NF-20-1-0226)  and National Science Foundation (grant\# CMMI-1554429) awarded to B.C.}

\showacknow 
\showfunding
\bibliography{Arxiv_main.bib}

\begin{thebibliography}{10}

\bibitem{dudley2002biomechanics}
R Dudley, {\em The biomechanics of insect flight: form, function, evolution}.
\newblock (Princeton University Press), (2002).

\bibitem{fry2003aerodynamics}
SN Fry, R Sayaman, MH Dickinson, The aerodynamics of free-flight maneuvers in
  drosophila.
\newblock {\em\protect\JournalTitle{Science}} \textbf{300}, 495--498 (2003).

\bibitem{shyy2013introduction}
W Shyy, H Aono, Ck Kang, H Liu, {\em An introduction to flapping wing
  aerodynamics}.
\newblock (Cambridge University Press) Vol.{}~37, (2013).

\bibitem{dickinson2005initiation}
MH Dickinson, The initiation and control of rapid flight maneuvers in fruit
  flies.
\newblock {\em\protect\JournalTitle{Integrative and comparative biology}}
  \textbf{45}, 274--281 (2005).

\bibitem{cheng2011translational}
B Cheng, X Deng, Translational and rotational damping of flapping flight and
  its dynamics and stability at hovering.
\newblock {\em\protect\JournalTitle{IEEE Transactions on Robotics}}
  \textbf{27}, 849--864 (2011).

\bibitem{sun2014insect}
M Sun, Insect flight dynamics: stability and control.
\newblock {\em\protect\JournalTitle{Reviews of Modern Physics}} \textbf{86},
  615 (2014).

\bibitem{cheng2016flight}
B Cheng, et~al., Flight mechanics and control of escape manoeuvres in
  hummingbirds. i. flight kinematics.
\newblock {\em\protect\JournalTitle{Journal of Experimental Biology}}
  \textbf{219}, 3518--3531 (2016).

\bibitem{bomphrey2017smart}
RJ Bomphrey, T Nakata, N Phillips, SM Walker, Smart wing rotation and
  trailing-edge vortices enable high frequency mosquito flight.
\newblock {\em\protect\JournalTitle{Nature}} \textbf{544}, 92--95 (2017).

\bibitem{taha2020vibrational}
HE Taha, M Kiani, TL Hedrick, JS Greeter, Vibrational control: A hidden
  stabilization mechanism in insect flight.
\newblock {\em\protect\JournalTitle{Science Robotics}} \textbf{5},
  eabb1502--eabb1502 (2020).

\bibitem{wood2008first}
RJ Wood, The first takeoff of a biologically inspired at-scale robotic insect.
\newblock {\em\protect\JournalTitle{IEEE transactions on robotics}}
  \textbf{24}, 341--347 (2008).

\bibitem{roll2015electromagnetic}
JA Roll, B Cheng, X Deng, An electromagnetic actuator for high-frequency
  flapping-wing microair vehicles.
\newblock {\em\protect\JournalTitle{IEEE Transactions on Robotics}}
  \textbf{31}, 400--414 (2015).

\bibitem{jafferis2019untethered}
NT Jafferis, EF Helbling, M Karpelson, RJ Wood, Untethered flight of an
  insect-sized flapping-wing microscale aerial vehicle.
\newblock {\em\protect\JournalTitle{Nature}} \textbf{570}, 491--495 (2019).

\bibitem{zhang2017design}
J Zhang, F Fei, Z Tu, X Deng, Design optimization and system integration of
  robotic hummingbird in {\em 2017 IEEE International Conference on Robotics
  and Automation (ICRA)}.
\newblock (IEEE), pp. 5422--5428 (2017).

\bibitem{karasek2018tailless}
M Kar{\'a}sek, FT Muijres, C De~Wagter, BD Remes, GC de~Croon, A tailless
  aerial robotic flapper reveals that flies use torque coupling in rapid banked
  turns.
\newblock {\em\protect\JournalTitle{Science}} \textbf{361}, 1089--1094 (2018).

\bibitem{liu1998numerical}
H Liu, K Kawachi, A numerical study of insect flight.
\newblock {\em\protect\JournalTitle{Journal of computational physics}}
  \textbf{146}, 124--156 (1998).

\bibitem{dickinson1993unsteady}
MH Dickinson, KG Gotz, Unsteady aerodynamic performance of model wings at low
  reynolds numbers.
\newblock {\em\protect\JournalTitle{Journal of experimental biology}}
  \textbf{174}, 45--64 (1993).

\bibitem{sun2002unsteady}
M Sun, J Tang, Unsteady aerodynamic force generation by a model fruit fly wing
  in flapping motion.
\newblock {\em\protect\JournalTitle{Journal of experimental biology}}
  \textbf{205}, 55--70 (2002).

\bibitem{young2009details}
J Young, SM Walker, RJ Bomphrey, GK Taylor, AL Thomas, Details of insect wing
  design and deformation enhance aerodynamic function and flight efficiency.
\newblock {\em\protect\JournalTitle{Science}} \textbf{325}, 1549--1552 (2009).

\bibitem{shyy2010recent}
W Shyy, et~al., Recent progress in flapping wing aerodynamics and
  aeroelasticity.
\newblock {\em\protect\JournalTitle{Progress in Aerospace Sciences}}
  \textbf{46}, 284--327 (2010).

\bibitem{song2014three}
J Song, H Luo, TL Hedrick, Three-dimensional flow and lift characteristics of a
  hovering ruby-throated hummingbird.
\newblock {\em\protect\JournalTitle{Journal of The Royal Society Interface}}
  \textbf{11}, 20140541 (2014).

\bibitem{bayiz2018real}
YE Bayiz, et~al., Real-time learning of efficient lift generation on a
  dynamically scaled flapping wing using policy search in {\em 2018 IEEE
  International Conference on Robotics and Automation, ICRA 2018}.
\newblock (IEEE), pp. 5519--5525 (2018).

\bibitem{bayiz2019experimental}
YE Bayiz, SJ Hsu, AN Aguiles, Y Shade-Alexander, B Cheng, Experimental learning
  of a lift-maximizing central pattern generator for a flapping robotic wing in
  {\em 2019 International Conference on Robotics and Automation (ICRA)}.
\newblock (IEEE), pp. 1997--2003 (2019).

\bibitem{minotti2002unsteady}
F Minotti, Unsteady two-dimensional theory of a flapping wing.
\newblock {\em\protect\JournalTitle{Physical Review E}} \textbf{66}, 051907
  (2002).

\bibitem{sane2002aerodynamic}
SP Sane, MH Dickinson, The aerodynamic effects of wing rotation and a revised
  quasi-steady model of flapping flight.
\newblock {\em\protect\JournalTitle{Journal of experimental biology}}
  \textbf{205}, 1087--1096 (2002).

\bibitem{ansari2006non}
S Ansari, R {\.Z}bikowski, K Knowles, Non-linear unsteady aerodynamic model for
  insect-like flapping wings in the hover. part 1: methodology and analysis.
\newblock {\em\protect\JournalTitle{Proceedings of the Institution of
  Mechanical Engineers, Part G: Journal of Aerospace Engineering}}
  \textbf{220}, 61--83 (2006).

\bibitem{berman2007energy}
GJ Berman, ZJ Wang, Energy-minimizing kinematics in hovering insect flight.
\newblock {\em\protect\JournalTitle{Journal of Fluid Mechanics}} \textbf{582},
  153 (2007).

\bibitem{whitney2010aeromechanics}
JP Whitney, RJ Wood, Aeromechanics of passive rotation in flapping flight.
\newblock {\em\protect\JournalTitle{Journal of fluid mechanics}} \textbf{660},
  197 (2010).

\bibitem{lee2016quasi}
Y Lee, KB Lua, T Lim, K Yeo, A quasi-steady aerodynamic model for flapping
  flight with improved adaptability.
\newblock {\em\protect\JournalTitle{Bioinspiration \& biomimetics}}
  \textbf{11}, 036005 (2016).

\bibitem{wolpert1995internal}
DM Wolpert, Z Ghahramani, MI Jordan, An internal model for sensorimotor
  integration.
\newblock {\em\protect\JournalTitle{Science}} \textbf{269}, 1880--1882 (1995).

\bibitem{wolpert2000computational}
DM Wolpert, Z Ghahramani, Computational principles of movement neuroscience.
\newblock {\em\protect\JournalTitle{Nature neuroscience}} \textbf{3},
  1212--1217 (2000).

\bibitem{shadmehr2012biological}
R Shadmehr, S Mussa-Ivaldi, {\em Biological learning and control: how the brain
  builds representations, predicts events, and makes decisions}.
\newblock (Mit Press), (2012).

\bibitem{doerr2018probabilistic}
A Doerr, et~al., Probabilistic recurrent state-space models in {\em Proceedings
  of the 35th International Conference on Machine Learning}.
\newblock pp. 1280--1289 (2018).

\bibitem{lua2016aerodynamic}
KB Lua, Y Lee, T Lim, K Yeo, Aerodynamic effects of elevating motion on
  hovering rigid hawkmothlike wings.
\newblock {\em\protect\JournalTitle{AIAA Journal}} \textbf{54}, 2247--2264
  (2016).

\bibitem{weis1973quick}
T Weis-Fogh, Quick estimates of flight fitness in hovering animals, including
  novel mechanisms for lift production.
\newblock {\em\protect\JournalTitle{Journal of experimental Biology}}
  \textbf{59}, 169--230 (1973).

\bibitem{ellington1996leading}
CP Ellington, C Van Den~Berg, AP Willmott, AL Thomas, Leading-edge vortices in
  insect flight.
\newblock {\em\protect\JournalTitle{Nature}} \textbf{384}, 626--630 (1996).

\bibitem{dickinson1999wing}
MH Dickinson, FO Lehmann, SP Sane, Wing rotation and the aerodynamic basis of
  insect flight.
\newblock {\em\protect\JournalTitle{Science}} \textbf{284}, 1954--1960 (1999).

\bibitem{nakata2015cfd}
T Nakata, H Liu, RJ Bomphrey, A cfd-informed quasi-steady model of
  flapping-wing aerodynamics.
\newblock {\em\protect\JournalTitle{Journal of fluid mechanics}} \textbf{783},
  323--343 (2015).

\bibitem{pohly2018quasi}
JA Pohly, JL Salmon, JE Bluman, K Nedunchezian, Ck Kang, Quasi-steady versus
  navier--stokes solutions of flapping wing aerodynamics.
\newblock {\em\protect\JournalTitle{Fluids}} \textbf{3}, 81 (2018).

\bibitem{taha2014state}
HE Taha, MR Hajj, PS Beran, State-space representation of the unsteady
  aerodynamics of flapping flight.
\newblock {\em\protect\JournalTitle{Aerospace Science and Technology}}
  \textbf{34}, 1--11 (2014).

\bibitem{sane2003aerodynamics}
SP Sane, The aerodynamics of insect flight.
\newblock {\em\protect\JournalTitle{Journal of experimental biology}}
  \textbf{206}, 4191--4208 (2003).

\bibitem{durbin2012time}
J Durbin, SJ Koopman, {\em Time series analysis by state space methods}.
\newblock (Oxford university press), (2012).

\bibitem{taha2021state}
HE Taha, AS Rezaei, State space modeling of viscous unsteady aerodynamic loads
  in {\em AIAA Scitech 2021 Forum}.
\newblock p. 1830 (2021).

\bibitem{wang2007gaussian}
JM Wang, DJ Fleet, A Hertzmann, Gaussian process dynamical models for human
  motion.
\newblock {\em\protect\JournalTitle{IEEE transactions on pattern analysis and
  machine intelligence}} \textbf{30}, 283--298 (2007).

\bibitem{deisenroth2012expectation}
M Deisenroth, S Mohamed, Expectation propagation in gaussian process dynamical
  systems in {\em Advances in Neural Information Processing Systems}.
\newblock pp. 2609--2617 (2012).

\bibitem{frigola2013bayesian}
R Frigola, F Lindsten, TB Sch{\"o}n, CE Rasmussen, Bayesian inference and
  learning in gaussian process state-space models with particle mcmc in {\em
  Advances in Neural Information Processing Systems}.
\newblock pp. 3156--3164 (2013).

\bibitem{frigola2014variational}
R Frigola, Y Chen, CE Rasmussen, Variational gaussian process state-space
  models in {\em Advances in neural information processing systems}.
\newblock pp. 3680--3688 (2014).

\bibitem{svensson2016computationally}
A Svensson, A Solin, S S{\"a}rkk{\"a}, T Sch{\"o}n, Computationally efficient
  bayesian learning of gaussian process state space models in {\em Artificial
  Intelligence and Statistics}.
\newblock pp. 213--221 (2016).

\bibitem{eleftheriadis2017identification}
S Eleftheriadis, T Nicholson, M Deisenroth, J Hensman, Identification of
  gaussian process state space models in {\em Advances in neural information
  processing systems}.
\newblock pp. 5309--5319 (2017).

\bibitem{williams2006gaussian}
CK Williams, CE Rasmussen, {\em Gaussian processes for machine learning}.
\newblock (MIT press Cambridge, MA) Vol.{}~2, (2006).

\bibitem{quinonero2005unifying}
J Qui{\~n}onero-Candela, CE Rasmussen, A unifying view of sparse approximate
  gaussian process regression.
\newblock {\em\protect\JournalTitle{Journal of Machine Learning Research}}
  \textbf{6}, 1939--1959 (2005).

\bibitem{snelson2006sparse}
E Snelson, Z Ghahramani, Sparse gaussian processes using pseudo-inputs in {\em
  Advances in neural information processing systems}.
\newblock pp. 1257--1264 (2006).

\bibitem{blei2017variational}
DM Blei, A Kucukelbir, JD McAuliffe, Variational inference: A review for
  statisticians.
\newblock {\em\protect\JournalTitle{Journal of the American statistical
  Association}} \textbf{112}, 859--877 (2017).

\bibitem{chen2018unsteady}
L Chen, J Wu, C Zhou, SJ Hsu, B Cheng, Unsteady aerodynamics of a
  pitching-flapping-perturbed revolving wing at low reynolds number.
\newblock {\em\protect\JournalTitle{Physics of Fluids}} \textbf{30}, 051903
  (2018).

\bibitem{friedman2001elements}
J Friedman, T Hastie, R Tibshirani, {\em The elements of statistical learning}.
\newblock (Springer series in statistics New York) Vol.{}~1, (2001).

\bibitem{zheng2020kinematic}
H Zheng, F Xie, T Ji, Y Zheng, Kinematic parameter optimization of a flapping
  ellipsoid wing based on the data-informed self-adaptive quasi-steady model.
\newblock {\em\protect\JournalTitle{Physics of Fluids}} \textbf{32}, 041904
  (2020).

\bibitem{chin2016flapping}
DD Chin, D Lentink, Flapping wing aerodynamics: from insects to vertebrates.
\newblock {\em\protect\JournalTitle{Journal of Experimental Biology}}
  \textbf{219}, 920--932 (2016).

\bibitem{birch2003influence}
JM Birch, MH Dickinson, The influence of wing--wake interactions on the
  production of aerodynamic forces in flapping flight.
\newblock {\em\protect\JournalTitle{Journal of experimental biology}}
  \textbf{206}, 2257--2272 (2003).

\bibitem{bowman1997applied}
AW Bowman, A Azzalini, {\em Applied smoothing techniques for data analysis: the
  kernel approach with S-Plus illustrations}.
\newblock (OUP Oxford) Vol.{}~18, (1997).

\bibitem{box2015time}
GE Box, GM Jenkins, GC Reinsel, GM Ljung, {\em Time series analysis:
  forecasting and control}.
\newblock (John Wiley \& Sons), (2015).

\bibitem{wu1998post}
JZ Wu, XY Lu, AG Denny, M Fan, JM Wu, Post-stall flow control on an airfoil by
  local unsteady forcing.
\newblock {\em\protect\JournalTitle{Journal of Fluid Mechanics}} \textbf{371},
  21--58 (1998).

\bibitem{amitay2001aerodynamic}
M Amitay, DR Smith, V Kibens, DE Parekh, A Glezer, Aerodynamic flow control
  over an unconventional airfoil using synthetic jet actuators.
\newblock {\em\protect\JournalTitle{AIAA journal}} \textbf{39}, 361--370
  (2001).

\bibitem{brunton2015closed}
SL Brunton, BR Noack, Closed-loop turbulence control: Progress and challenges.
\newblock {\em\protect\JournalTitle{Applied Mechanics Reviews}} \textbf{67}
  (2015).

\bibitem{ijspeert2008central}
AJ Ijspeert, Central pattern generators for locomotion control in animals and
  robots: a review.
\newblock {\em\protect\JournalTitle{Neural networks}} \textbf{21}, 642--653
  (2008).

\bibitem{manoonpong2013neural}
P Manoonpong, U Parlitz, F W{\"o}rg{\"o}tter, Neural control and adaptive
  neural forward models for insect-like, energy-efficient, and adaptable
  locomotion of walking machines.
\newblock {\em\protect\JournalTitle{Frontiers in neural circuits}} \textbf{7},
  12 (2013).

\bibitem{kang2018minimum}
L Kang, L Liu, W Su, J Wu, Minimum-domain impulse theory for unsteady
  aerodynamic force.
\newblock {\em\protect\JournalTitle{Physics of Fluids}} \textbf{30}, 016107
  (2018).

\bibitem{cheng2014three}
B Cheng, J Roll, Y Liu, DR Troolin, X Deng, Three-dimensional vortex wake
  structure of flapping wings in hovering flight.
\newblock {\em\protect\JournalTitle{Journal of The Royal Society Interface}}
  \textbf{11}, 20130984 (2014).

\end{thebibliography}

\end{document}